\documentclass[12pt]{article}
\usepackage[english,german,french,polish]{babel}
\usepackage[T1]{fontenc}
\usepackage{amsfonts}

\begin{document}

\selectlanguage{english}

\baselineskip 0.76cm
\topmargin -0.4in
\oddsidemargin -0.1in

\let\ni=\noindent

\renewcommand{\thefootnote}{\fnsymbol{footnote}}

\newcommand{\SM}{Standard Model }

\pagestyle {plain}

\setcounter{page}{1}



~~~~~~
\pagestyle{empty}

\begin{flushright}
IFT-- 07/14
\end{flushright}

\vspace{0.4cm}

{\large\centerline{\bf Photonic portal to the sterile world of cold dark matter{\footnote{Work supported in part by the Polish Ministry of Science and Higher Education  under the grant 1\,PO3B 099 29 (2005-2007). }}}}

\vspace{0.5cm}

{\centerline {\sc Wojciech Kr\'{o}likowski}}

\vspace{0.3cm}

{\centerline {\it Institute of Theoretical Physics, Warsaw University}}

{\centerline {\it Ho\.{z}a 69,~~PL--00--681 Warszawa, ~Poland}}

\vspace{0.6cm}

{\centerline{\bf Abstract}}

\vspace{0.2cm}

\begin{small}

\begin{quotation}

 We assume that the cold dark matter consists of spin-1/2 and spin-0 particles described by a bispinor field
$\psi$ and a scalar field $\varphi$, sterile from all \SM charges (in contrast, neutralinos, supersymmetric  candidates for cold dark matter, are not sterile from weak \SM charges). We propose, however, that such a sterile world can contact with our \SM world not only through gravity but also through a portal provided by
photons coupled to sterile particles by means of two very weak effective interactions $-(f/M^2)\varphi F^{\mu  \nu}\varphi F_{\mu\nu}$ and $-(f'/M^2) (\bar{\psi}\sigma^{\mu\nu}\psi )\varphi F_{\mu\nu}$, where $M$ is a 
very large mass scale and $f$ and $f'$ are dimensionless coupling constants. Thus, in our picture, the 
electromagnetic field $F_{\mu\nu}$ -- as the only \SM field --$\!$ participates in both worlds, providing a 
non\-gravitational link between them (other than the popular supersymmetric weak interaction, active in the 
case of neutralinos). In consequence, there appears a tiny quasi-magnetic correction to the conventional 
electromagnetic current (described in Appendix A).

\vspace{0.2cm}
 
\ni PACS numbers: 14.80.-j , 04.50.+h , 95.35.+d 

\vspace{0.2cm}

\ni December 2007

\end{quotation}

\end{small}

\vfill\eject

\pagestyle {plain}

\setcounter{page}{1}

\vspace{0.4cm}

\ni {\bf 1. Introduction: content of dark matter}

\vspace{0.4cm}

All presently known fundamental fermions {\it i.e.},  leptons and quarks, carry the  \SM charges and so, are coupled to the \SM gauge bosons. Leptons differ from quarks by not participating in $SU(3)$ strong interactions, though both display $SU(2)\times U(1)$ electroweak interactions. Besides, leptons and quarks (as well as the \SM gauge and Higgs bosons) are coupled also to the gravitational field that, if successfully quantized, is represented by gravitons (collaborating, perhaps, with their dilaton-like and/or axion-like partners).
 
In this situation, one may ask a so-called good question, if in Nature there is a place for a sort of fundamental fermions interacting only gravitationally. Such fundamental fermions are by definition sterile from all \SM charges and do not mix with active neutrinos and so, they are blind to all \SM gauge interactions and do not cooscillate with active neutrinos. These sterile spin-1/2 particles will be called here {\it sterinos} (a shortening from sterilinos) to get a short name emphasizing both their sterility and their half-integer spin.
 
It is very natural to wonder, if just the sterinos --- rather than neutralinos of the supersymmetric extension of the \SM --- can be responsible for the fermion component of cosmic cold dark matter which dominates globally in all matter of our Universe and, at present, is one of the most important problems of today's particle physics and astrophysics [1].

The experimental search for physical effects caused by possible interactions of cold dark matter other than gravity (extended, perhaps, by the action of hypothetic spin-0 partners of gravitons) is presently in the centre of attention. This search can be classified either as direct detection experiments [2], where one hopes to observe recoils of nuclei scattered elastically from dark matter particles, or indirect experiments [3], where one tries to identify annihilation or decay products of dark matter as {\it e.g.} positrons possibly created by dark matter in the centre of our Galaxy and subsequently annihilated at rest with the emission of 511 keV line observed since 1970s.

Sterinos require a considerable mass in order to participate in the cold dark matter. One may speculate that in Nature there are fundamental scalar bosons, also sterile from all \SM charges [4], whose field --- like the neutral component of \SM Higgs boson field --- develops a nonzero vacuum expectation value. It will be convenient to use for these sterile scalar particles the name {\it sterons} (a shortening from sterilons).

The terms in the Lagrangian needed to generate the sterino Dirac mass may be taken in the form

\begin{equation}
-\bar{\psi}_{\rm sto} \,y\, \psi_{\rm sto} \varphi_{\rm stn} + \frac{1}{2}\mu^2 \varphi^2_{\rm stn} - \frac{1}{4}\lambda \varphi^4_{\rm stn} 
\end{equation}

\ni with

\begin{equation}
< \varphi_{\rm stn}>_{\rm vac} = \frac{\mu}{\sqrt{\lambda}} \neq 0
\end{equation}

\vspace{0.2cm}

\ni (in the tree approximation), where $y$ is an unknown Yukawa coupling constant (being a matrix in the case of sterinos developing more generations). The constants $\mu > 0$ and $\lambda > 0$ appearing in the steron potential $V(\varphi_{\rm stn}) = -(1/2)\mu^2 \varphi^2_{\rm stn} + (1/4)\lambda \varphi^4_{\rm stn}$ are also unknown. Then, the sterino Dirac mass becomes

\begin{equation}
m^{(D)}_{\rm sto} = y <\varphi_{\rm stn}>_{\rm vac} = y \frac{\mu}{\sqrt{\lambda}} \,.
\end{equation}

\vspace{0.2cm}

\ni The physical steron field is given by the difference $\varphi^{\rm (ph)}_{\rm stn} \equiv \varphi_{\rm stn} -  <\varphi_{\rm stn}>_{\rm vac}$. Then, the mass of physical sterons is generated as

\begin{equation} 
m_{\rm stn}  = \mu \sqrt{2} = \sqrt{2\lambda} <\varphi_{\rm stn}>_{\rm vac}\,,
\end{equation}

\vspace{0.4cm}

\ni since $\,V(<\varphi_{\rm stn}>_{\rm vac}\,+\,\varphi^{\rm (ph)}_{\rm stn}) \,=\, \mu^2 \varphi^{{\rm (ph)}\,2}_{\rm stn} \,+\, \mu\, \sqrt{\lambda}\, \varphi^{{\rm (ph)}\,3}_{\rm stn} \,+\, (1/4)\lambda\, \varphi^{{\rm (ph)}\,4}_{\rm stn} - (1/4)\mu^4/\lambda  $. The physical sterons may exist in Nature as the (probably unstable) boson component of cold dark matter.

Thus, the nonzero vacuum expectation value of the steron field $\varphi_{\rm stn}$ breaks spontaneously the scale symmetry of the sterile world consisting of sterinos and sterons. The righthanded neutrinos, as being sterile, should also belong to this world. {\it A priori}, they may be either different or identical with the righthanded components of sterinos. But, in both options their Majorana mass $m^{(R)}_{\nu}$, usually considered as very large, may be generated by the same vacuum expectation value of the steron field $\varphi_{\rm stn}$ which generates also the sterino Dirac mass $m^{(D)}_{\rm sto}$. Then, this expectation value breaks spontaneously once more the scale symmetry of the sterile world. Restricting oneself to a minimal picture of sterile world, one may try to imagine that the second option is true {\it i.e.}, the righthanded neutrino field is identical with the righthanded component of sterino field,

\begin{equation}
\nu_R \equiv \psi_{{\rm sto}\,R} \,.
\end{equation}

\ni Then, the combined fields $\nu = \nu_L + \nu_R$ and $\psi_{\rm sto} = \psi_{{\rm sto}\,L} + \psi_{{\rm sto}\,R}$ get the common righthanded component $\nu_R$. We contest this option in the footnote \dag.

\vspace{0.4cm}

\ni {\bf 2. Generations in dark matter}

\vspace{0.4cm}

The sterino Dirac mass (3) is the only existing kind of sterino mass, if sterinos are Dirac fermions. If, however, they are Majorana fermions, then one can define for them a more general mass term in the Lagrangian, namely


\begin{equation}
-\frac{1}{2}\left(\overline{{\psi_{{\rm sto}\,L}}}, (\overline{\psi_{{\rm sto}\,R})^c}\right)\, \left(\begin{array}{ll} m^{(L)}_{\rm sto} & m^{(D)}_{\rm sto} \\ m^{(D)\,T}_{\rm sto} & m^{(R)}_{\rm sto} \end{array}\right)\,\left(\begin{array}{c} (\psi_{{\rm sto}\,L})^c \\ \psi_{{\rm sto}\,R}\,  \end{array}\right) + h.c. \,,
\end{equation}

\vspace{0.4cm}

\ni where $\psi = \psi_L + \psi_R$, $\psi^c = C\bar\psi^T = -\beta C \psi^*$ and $\overline{\psi^c} = \psi^{c\,\dagger}\beta = - \psi^T C^{-1}\; (\dagger = * T)$. The Dirac and Majorana righthanded parts of this mass term are
 
\begin{equation}
- \overline{\psi_{\rm sto}}\,m^{(D)}_{\rm sto}\,\psi_{\rm sto}
\end{equation}

\vspace{-0.2cm}

\ni and

\vspace{-0.2cm}

\begin{equation}
\frac{1}{2}\left[(\psi_{{\rm sto}\,R})^T\, C^{-1} m^{(R)}_{\rm sto} \psi_{{\rm sto}\,R} + 
(\psi_{{\rm sto}\,R})^{T*}\, C m^{(R)}_{\rm sto} (\psi_{{\rm sto}\,R})^*\right] \,,
\end{equation}

\vspace{0.2cm}

\ni respectively. A similar form can be written down for the Majorana lefthanded part.

In Eqs. (6), (7) and (8), the masses $m^{(L)}_{\rm sto}, m^{(R)}_{\rm sto}$ and $m^{(D)}_{\rm sto}$ (as well as the Yukawa coupling constant $y$ in Eqs. (1) and (3)) are actually $2\times 2$ or $3\times 3$ matrices, when two or three generations of sterinos exist in Nature {\it i.e.}, when $\psi_{\rm sto}$ is a doublet or triplet of sterino fields{\footnote{In a series of papers, we have formulated an "intrinsic interpretation" of three lepton and quark generations, based on a generalization of Dirac's square root procedure [5]. In addition, we have found there that for hypothetic \SM fundamental spin-0 particles (Higgs bosons) as well as for hypothetic fundamental spin-1/2 and spin-0 particles, sterile from all \SM charges, there should exist two generations rather than one or three. Of course, in such an approach, the identity (5) cannot work for three generations of righthanded neutrinos $\nu_R$. However, a peculiar option still exists, where $\nu_R$ and $\psi_{{\rm sto}\,R}$ appear in two generations and so, the identity (5) may hold for two generations in spite of the fact that there are three generations for $\nu_L$.}}. If, in contrast to the footnote \dag, the identification (5) holds for three generations of $\nu_R$, then $m^{(R)}_{\rm sto} $ is identical with the $3\times 3$  Majorana mass matrix for righthanded neutrinos of three generations, $m_\nu^{(R)}$, involved in the general neutrino mass matrix in the Lagrangian 

\vspace{0.1cm}

\begin{equation}
-\frac{1}{2}\left(\overline{\nu_L}, \overline{(\nu_R)^c}\right)\, \left(\begin{array}{ll}0 & m^{(D)}_\nu \\ m^{(D)\,T}_\nu & m^{(R)}_\nu \end{array}\right)\,\left(\begin{array}{c} (\nu_L)^c \\ \nu_R\, \end{array}\right) + h.c. \,,
\end{equation}

\ni so that 

\begin{equation}
m^{(R)}_\nu \equiv  m^{(R)}_{\rm sto} \,.
\end{equation}

\vspace{0.1cm}

If in the mass term (6) $m^{(R)}_{\rm sto}$ dominates over $m^{(L)}_{\rm sto}$ and $m^{(D)}_{\rm sto}$, then two mass states, formed in the diagonalization procedure in Eq. (6), get approximately the masses 

\begin{equation}
m^{(L)}_{\rm sto} - m^{(D)}_{\rm sto}m^{(R)\,-1}_{\rm sto}m^{(D)\,T}_{\rm sto} \,,\, m^{(R)}_{\rm sto} + 
m^{(D)}_{\rm sto} m^{(R)\,-1}_{\rm sto}m^{(D)\,T}_{\rm sto} \simeq m^{(R)}_{\rm sto} \,,
\end{equation}

\vspace{0.1cm}

\ni analogical to those in the case of seesaw mechanism for neutrinos [6].

If, on the contrary, $m^{(D)}_{\rm sto}$ dominates over $m^{(L)}_{\rm sto}$ and $m^{(R)}_{\rm sto}$, then two sterino mass states, constructed for sterinos in the diagonlization procedure in Eq. (6), obtain approximately the masses

\begin{equation}
\mp m^{(D)}_{\rm sto} + \frac{1}{2}\left( m^{(L)}_{\rm sto} + m^{(R)}_{\rm sto}\right) \simeq 
\mp m^{(D)}_{\rm sto}\,,
\end{equation}

\vspace{0.1cm}

\ni similar to those in the case of pseudoDirac neutrinos. 

When $m^{(L)}_{\rm sto}$, $m^{(R)}_{\rm sto}$ and $m^{(D)}_{\rm sto}$ in Eqs. (11) and (12) are $2\times 2$ or $3\times 3$ sterino mass matrices, then these equations, resulting in the first step of digonalization in Eq. (6), still require the second step which leads eventually to two doublets or triplets of sterino mass corresponding to two doublets or triplets of sterino mass states.

\vspace{0.4cm}

\ni {\bf 3. Photonic portal to dark matter}

\vspace{0.4cm}

In our picture, therefore, the sterile world consists of sterinos and sterons and, in addition, righthanded neutrinos if the latter are different from righthanded sterinos. Such a sterile world contacts with our \SM world through the exchange of gravitons (collaborating with their hypothetic spin-0 partners) and, possibly, also by means of \SM Higgs bosons playing the role of a {\it Higgs portal} to the sterile world [4,7], if {\it Higgs bosons interact directly with sterons} through an additional term in the Lagrangian. This renders the physical scalar bosons of both sorts mixed. 

An alternative to this Higgs portal may be a {\it photonic portal} to the sterile world provided by the \SM electromagnetic field $F_{\mu\,\nu}$ acting as {\it the pair $F^{\mu\,\nu}\,F_{\mu\,\nu}$ interacting directly with the pair $\varphi_{\rm stn}\,\varphi_{\rm stn}$} of the steron field $\varphi_{\rm stn} $ through an additional term in the Lagrangian  

\begin{equation}
-\frac{f}{M^2}\, \varphi^2_{\rm stn} F^{\mu\,\nu} F_{\mu\,\nu} \,,
\end{equation}

\ni where $M$ denotes a very large mass scale and $f>0$ is an unknown dimensionless coupling constant{\footnote{It may be convenient to replace the constant $f$ in Eq. (13) by $f/4$ because of the normalization of scalar $F^{\mu\,\nu}\,F_{\mu\,\nu}$ in the Lagrangian (see Appendix A). }}. Here, $\varphi^2_{\rm stn} = \varphi^{({\rm ph})}_{\rm stn} +2(\mu/\sqrt{\lambda}) \varphi^{({\rm ph})\,2}_{\rm stn}  + \mu^2/\lambda$. Due to the very large $M$, the photonic portal is very narrow at low energies. Note that the bilinear form $\varphi_{\rm stn} F^{\mu\,\nu}$ in the effective interaction (13) of dimension six plays the role of an antisymmetric tensor current, coupled to itself.

One can speculate that --- on a more fundamental level --- the antisymmetric tensor current $\varphi_{\rm stn} F^{\mu\,\nu}$ is coupled to a very massive antisymmetric tensor field $A_{\mu\,\nu}$ of dimension one: $\propto \varphi_{\rm stn} F^{\mu\,\nu} A_{\mu\,\nu}$ (the field $F_{\mu\,\nu} = \partial_\mu A_\nu - \partial_\nu A_\mu $ is of dimension two). The new field $A_{\mu\,\nu}$ comprises two three-dimensional fields: vector $\vec{A}_E = (A_{01},\,A_{02},\,A_{03})$ and axial $\vec{A}_H = (A_{23},\,A_{31},\,A_{12})$, so it describes two kinds of sterile spin-1 particles with the parity -1 and +1, respectively, and with a very large mass $M$. The exchange of these bosons leads to the current$\times $current effective interaction (13), when momentum transfers through the field $A_{\mu\,\nu}$ are negligible in comparison with $M$. One may speculate further that the conventional tensor current formed of sterinos, $\bar\psi_{\rm sto} \sigma^{\mu\,\nu} \psi_{\rm sto}$, is also coupled to $A_{\mu\,\nu}:\; \propto (\bar\psi_{\rm sto} \sigma^{\mu\,\nu} \psi_{\rm sto}) A_{\mu\,\nu}$, leading to two extra effective interactions

\vspace{-0.1cm}

\begin{equation}
-\frac{f'}{M^2}\, (\bar\psi_{\rm sto} \sigma^{\mu\,\nu} \psi_{\rm sto} ) \varphi_{\rm stn} F_{\mu\,\nu}  \;\;\,,\;\;\, -\frac{f''}{M^2}\,(\bar\psi_{\rm sto} \sigma^{\mu\,\nu} \psi_{\rm sto} ) (\bar\psi_{\rm sto} \sigma_{\mu\,\nu} \psi_{\rm sto} ) 
\end{equation}

\ni with unknown dimensionless coupling constants $f'>0$ and $f''>0$, when momentum transfers {\it via} the field $A_{\mu\,\nu}$ can be neglected {\it versus} $M$. If the field $A_{\mu\,\nu}$ is coupled universally to the bilinear form $\varphi_{\rm stn} F^{\mu\,\nu} + \zeta  \bar\psi_{\rm sto} \sigma^{\mu\,\nu} \psi_{\rm sto}$ playing the role of total antisymmetric tensor current where $\zeta > 0$ is an unknown constant: $\propto (\varphi_{\rm stn} F^{\mu\,\nu} + \zeta  \bar\psi_{\rm sto} \sigma^{\mu\,\nu} \psi_{\rm sto}) A_{\mu\,\nu}$, then in the case of vanishing momentum transfers through the field $A_{\mu\,\nu}$ the universal effective interaction 

\vspace{-0.1cm}

\begin{equation}
-\frac{f}{M^2}\, (\varphi_{\rm stn} F^{\mu\,\nu} + \zeta  \bar\psi_{\rm sto} \sigma^{\mu\,\nu} \psi_{\rm sto} ) (\varphi_{\rm stn} F_{\mu\,\nu} + \zeta \bar\psi_{\rm sto} \sigma_{\mu\,\nu} \psi_{\rm sto} ) 
\end{equation}

\ni follows. This implies that

\vspace{-0.1cm}

\begin{equation}
f : f' : f'' = 1 : 2\zeta : \zeta^2\,,
\end{equation}

\ni when compared with Eqs. (13) and (14). Eventually, the pair $A^{\mu\,\nu}A_{\mu\,\nu}$ may be coupled to the pair $\varphi_{\rm stn} \varphi_{\rm stn}:\; \propto \varphi^2_{\rm stn} A^{\mu\,\nu}A_{\mu\,\nu}$, generating the mass $M \propto <\varphi_{\rm stn}>_{\rm vac}$ for the field $A_{\mu\,\nu}$.

\vspace{0.4cm}

\ni {\bf 4. Examples of annihilation and  decay of dark matter}

\vspace{0.4cm}

The interaction (13) and the first interaction (14) might be considered as responsible for the phenomenon of low energy positrons, boldly presumed by C.~Boehm and collaborators [8] to be created in the centre of our Galaxy in  process of dark matter annihilation and subsequently annihilated at rest with the emission of 511 keV line observed since 1970s (in this case, the cold dark matter ought to be  considerably light).

If steron and sterino masses are appropriate, the hypothetic process initiated in our case by the interaction (13) and the first interaction (14) should run as follows:

\begin{equation}
\left.\begin{array}{c} ({\rm steron})({\rm steron}) \\ {\rm or} \\ ({\rm steron}) \end{array}\right\} \rightarrow \gamma_{\rm virt}\, \gamma \rightarrow e^+ \, e^- \,\gamma\; 
\end{equation}

\ni and

\begin{equation}
({\rm antisterino})({\rm sterino}) \rightarrow \left\{ \begin{array}{c} \gamma_{\rm virt}\, ({\rm steron}) \rightarrow e^+\, e^-({\rm steron}) \\ {\rm or} \\ \gamma_{\rm virt} \rightarrow e^+ \, e^- \end{array} \right.
\,,
\end{equation}

\vspace{0.2cm}

\ni respectively, and subsequently

\begin{equation}
 e^+\, e^- \rightarrow \gamma_{511}\,\gamma_{511} \;\;{\rm or}\;\;  e^+\, e^- \rightarrow ({\rm positronium}) 
\,\gamma_{\rm soft} \rightarrow \gamma_{511}\gamma_{511}\gamma_{\rm soft} \,.
\end{equation}

\ni Here, the created positron $e^+$ is annihilated afterwards at rest in pair with an electron $e^-$, other than the primarily created $e^-$. In the first processes (17) and (18) steron is obviously the physical steron, while in the second processes (17) and (18) one steron field has been replaced by its vacuum expectation value, 
$<{\rm steron}>_{\rm vac}$, the remaining steron being physical. 

Making use of our effective interactions (13) and (14) (as well as the \SM electromagnetic coupling $ - e 
\bar{\psi}_e \gamma^\mu \psi_e A_\mu$ for electrons), we can calculate the probabilities for processes (17) and (18), respectively.

For instance, the total cross-section for the first process (17) (the annihilation of a physical steron pair into an electron-positron pair and a photon) multiplied by the steron velocity is given in the steron centre-of-mass frame as follows:

\begin{equation} 
\sigma({\rm stn\; stn}\rightarrow{e^+ e^- \gamma}) v_{\rm stn} =\frac{1}{(2\pi)^3} \left( \frac{e\,f}{M^2} \right)^2 \frac{16}{3} \omega^2_{\rm stn} \,,
\end{equation}

\ni if the electron mass $m_e$ can be neglected. Here, $m_{\rm stn}$ and $\omega_{\rm stn}= 
\sqrt{\vec{p}\,^2_{\rm stn} + m^2_{\rm stn}}$ is the steron mass and the steron energy, respectively, while the steron velocity $v_{\rm stn} =|\vec{p}_{\rm stn}|/\omega_{\rm stn} = \sqrt{1-m^2_{\rm stn}/\omega^2_{\rm stn}}$ may be replaced by an average value of $v_{\rm stn}$, implying (through Eq. (20)) an average cross-section $\sigma({\rm stn\,stn} \rightarrow e^+\, e^-\, \gamma) $ dependent on an average energy squared $\omega^2_{\rm stn}$. In the centre-of-mass frame, the relative velocity of colliding sterons is $2v_{\rm stn}$.

Similarly, the total rate for the second process (17) (the decay of a physical steron into an electron-positron pair and a photon) is at rest equal to 

\vspace{-0.2cm}

\begin{equation}
\Gamma({\rm stn}\rightarrow{e^+ e^- \gamma}) = \frac{1}{(2\pi)^3} \left(\frac{e\,f <\varphi_{\rm stn}>_{\rm vac} }{M^2}\right)^2 \,\frac{4}{3}\,\omega^3_{\rm stn} = \frac{1}{6\pi^3} \left(\frac{e\,f/\sqrt{2\lambda}}{M^2}\right)^2 \,m^5_{\rm stn}\,,
\end{equation}

\ni if the electron mass is negligible. Here, at rest $\omega_{\rm stn} = m_{\rm stn}$, while $<\varphi_{\rm stn}>_{\rm vac}= m_{\rm stn}/\sqrt{2\lambda}$ (Eq. (4)) denotes the vacuum expectation value of the steron field. With $(1/4) G_{\rm eff}/\sqrt2 \equiv (e f/\sqrt{2\lambda})/M^2$ (see the footnote $\ddagger$), the steron rate (21) can be rewritten as $\Gamma({\rm stn}\rightarrow{e^+ e^- \gamma})  = G^2_{\rm eff} m^5_{\rm stn}/(192 \pi^3)$, where the rhs reminds formally of the total rate for  muon decay.

However, the simplest annihilation channel for a physical steron pair and decay channel for a physical steron is

\vspace{-0.2cm}

\begin{equation} 
({\rm steron})({\rm steron}) \rightarrow \gamma\, \gamma
\end{equation}

\vspace{-0.3cm}

\ni and

\vspace{-0.3cm}

\begin{equation} 
({\rm steron}) \rightarrow \gamma\, \gamma\,,
\end{equation}

\ni respectively. In this case, one gets respectively the following formulae for the total cross-section multiplied by the steron velocity:

\vspace{-0.2cm}

\begin{equation}
\sigma({\rm stn\; stn}\rightarrow{\gamma \gamma})  v_{\rm stn} = \frac{1}{2\pi} \left( \frac{f}{M^2} \right)^2 \omega^2_{\rm stn} 
\end{equation}

\ni in the steron centre-of-mass frame, and the total rate: 

\begin{equation}
\Gamma({\rm stn}\rightarrow{\gamma \gamma}) = \frac{1}{2\pi} \left(
\frac{f<\varphi_{\rm stn}>_{\rm vac} }{M^2}\right)^2\,\frac{1}{4}\,\omega^3_{\rm stn} = \frac{1}{8\pi} \left( \frac{f/\sqrt{2\lambda} }{M^2} \right)^2 \,m^5_{\rm stn}
\end{equation}

\ni at rest (where $\omega_{\rm stn} = m_{\rm stn}$). 

The large number of produced photons provided by the steron mechanism (17) and by the annihilation (22) and decay (23) may be inconsistent with observations, when the mechanism is fitted to the required positron production. In contrast, the sterino mechanism (18), when considered for the positron production, is in a better situation, since in this case the simplest annihilation channel contains only one photon,

\begin{equation}
({\rm antisterino})({\rm sterino}) \rightarrow \gamma\, ({\rm steron}) 
\end{equation}

\ni (if masses allow) and, first of all, the single-sterino state is stable under interactions of our photonic portal  giving  --- contrarily to the unstable single-steron state --- no additional photons. Moreover, the channel (26) may be energetically closed for the steron mass $m_{\rm stn}$ large enough, what may restrict further the number of photons produced in sterino-antisterino annihilation (for the total cross-section in the channel (26) see the end of Appendix B; also the elastic scattering of electrons on sterinos is calculated in some detail in this Appendix).
 
The steron decay channels open through the photonic portal cause that only sterinos remain as our (nonchiral) candidates for  stable dark matter (at least, sterinos of the lowest generation).

\vspace{0.4cm}

\ni {\bf 5. Final remarks}

\vspace{0.4cm}

We would like to stress finally that it is still possible that --- in reality --- the direct coupling exists neither between Higgs bosons and sterons nor between photons, sterons and sterino-antisterino pairs, so that in the \SM world there is no Higgs nor photonic portal to the sterile world. Then, only gravitons and, perhaps, also dilaton-like scalar and/or axion-like pseudoscalar partners of gravitons [9, 10] can mediate between both worlds as well as within the sterile world itself (of course, they can mediate also within the \SM world itself, but there they are dominated at the atomic scale by \SM media). Such a puristic picture still may explain the fundamental phenomenon of cold dark matter and, perhaps, provide a gravitational interpretation of the equally fundamental phenomenon of dark energy.

\vfill\eject

\vspace{0.3cm}

{\centerline{\bf Appendix A}} 

\vspace{0.3cm}

{\centerline{\it Quasi-electromagnetic current induced by dark matter}} 

\vspace{0.3cm}

In connection with the footnote $\ddagger$, it is worthwhile to observe that the hypothetic effective interaction (13) implies the free electromagnetic term in the Lagrangian being supplemented to the form 

$$
-\frac{1}{4} F^{\mu\,\nu} F_{\mu\,\nu} \rightarrow -\frac{1}{4}F^{\mu\,\nu} F_{\mu\,\nu}\left(1 + \frac{4f}{M^2} \varphi^2_{\rm stn} \right)\;. 
\eqno{\rm (A1)}
$$

\ni Then, the electromagnetic Lagrangian, supplemented as well by the hypothetic effective interaction (14) (with $f' = 2\zeta f $), is

$$
{\cal{ L }} = -\frac{1}{4} F^{\mu\,\nu} F_{\mu\,\nu} \left(1+ \frac{4f}{M^2} \varphi^2_{\rm stn} \right) - \frac{2\zeta f}{M^2}\bar\psi_{\rm sto} \sigma^{\mu\,\nu} \psi_{\rm sto} F_{\mu\,\nu} \varphi_{\rm stn} - j^\mu A_\mu\;. 
\eqno{\rm (A2)}
$$

\ni This leads to the following electromagnetic field equation:

$$
\partial_\nu F^{\mu\,\nu} = -\left(j^\mu + \delta j^\mu \right) 
\eqno{\rm (A3)}
$$

\ni with $F_{\mu \nu} =\partial_\mu A_\nu - \partial_\nu A_\mu$, where the additional current

$$
\delta j^\mu \equiv \frac{4f}{M^2} \partial_\nu\left[\varphi_{\rm stn} \left(\varphi_{\rm stn} F^{\mu\,\nu}   + \zeta \bar\psi_{\rm sto} \sigma^{\mu\,\nu} \psi_{\rm sto}\right)\right] 
\eqno{\rm (A4)}
$$

\ni is a quasi-magnetic correction induced by the photonic effective interactions (13) and (14) of the cold dark matter. Here, evidently,

$$
\partial_\mu \left(j^\mu + \delta j^\mu \right) =0
\eqno{\rm (A5)}
$$

\ni with the additional part $\partial_\mu \delta j^\mu $ being zero identically like the anomalous magnetic part of the conventional $\partial_\mu j^\mu $ in an effective presentation. Thus, $\partial_\mu j^\mu $ is zero dynamically for the conventional electromagnetic current $ j^\mu $, while $\delta j^\mu $ {\it supplements effectively the anomalous magnetic part of the conventional} $ j^\mu $.

\vfill\eject

\vspace{0.3cm}

{\centerline{\bf Appendix B}} 

\vspace{0.3cm}

{\centerline{\it Scattering of electrons on dark matter through photonic portal}} 

\vspace{0.3cm}

Due to our photonic portal the \SM world can interact quasi-electromagnetically with the cold dark matter. Consider for illustration the elastic scattering of electrons on sterinos,  making use of the first interaction (14) with $f' = 2\zeta f$ (and the \SM electromagnetic coupling $- e\,\bar{\psi}_e \gamma^\mu \psi_e A_\mu $ for electrons with  $e = -|e|$). The corresponding $S$ matrix element is

\begin{eqnarray*}
S_{f i} & = & \frac{2 e \zeta f <\varphi_{\rm stn}>_{\rm vac} }{M^2} \left[\frac{1}{(2\pi)^{12}}\, 
\frac{m^2_e m^2_{\rm sto}}{E'_e E_e\, E'_{\rm sto} E_{\rm sto} }\right]^{1/2} \left(2\pi \right)^4 \delta^4 \left( p'_e \!+\! p'_{\rm sto} \!-\! p_e \!-\! p_{\rm sto}\right) \\
 & \times & \bar{u}'_e(p'_e) \frac{1}{i} \left(k_\mu \gamma_\nu - k_\nu \gamma_\mu \right) u_e(p_e) 
\frac{-i}{k^2} \bar{u}'_{\rm sto}(p'_{\rm sto}) \sigma^{\mu\,\nu} u_{\rm sto}(p_{\rm sto}) 
\end{eqnarray*}

\vspace{-1.56cm}

\begin{flushright}
({\rm B}1)
\end{flushright}

\vspace{0.2cm}

\ni with the obvious notation. Here,

$$
k =  p_e  - p'_e = p'_{\rm sto} - p_{\rm sto}\,.
\eqno{\rm (B2)}
$$

\ni The factor appearing in the second line of Eq. (B1) can be obviously rewritten as 

$$
-2i \bar{u}'_e(p'_e) \gamma_\nu u_e(p_e)\frac{1}{k^2}\bar{u}'_{\rm sto}(p'_{\rm sto}) \left[2m_{\rm sto}\gamma^\nu - (p'_{\rm sto} + p_{\rm sto})^\nu  \right]  u_{\rm sto}(p_{\rm sto}) \,,
\eqno{\rm (B3)}
$$

\ni when the Gordon identity

$$
\bar{u}'(p') \gamma^\nu u(p) = \bar{u}'(p') \left[\frac{(p'+p)^\nu}{2m} + \frac{i \sigma^{\nu\mu}(p'-p)_\mu} {2m} \right] u(p)
\eqno{\rm (B4)}
$$

\ni is applied.

Hence, we can calculate the fully differential cross-section

\begin{eqnarray*}
\frac{d^6 \sigma}{d^3 \vec{p}\,'_e d^3 \vec{p}\,'_{\rm sto}} \!\!\!& \!\!=\!\! &  \frac{(2\pi)^6}{v_{\rm flux}} \,\frac{1}{4} \sum_{u'_e u'_{\rm sto}}\,\sum_{u_e u_{\rm sto}}\,\frac{|S_{f i}|^2}{(2\pi)^4 \delta^4(0)} \\
&\!\!=\!\! & \frac{1}{v_{\rm flux}}\left( \frac{2e \zeta f <\varphi_{\rm stn}>_{\rm vac}}{2\pi M^2}\right)^2 \, 
\frac{m^2_e m^2_{\rm sto}}{E'_e E_e E'_{\rm sto} E_{\rm sto}} \delta^4 \left(p'_e \!+\! p'_{\rm sto} \!-\!p_e \!-\!p_{\rm sto} \right) \\
& \!\!\times\!\! & \!\!\!\! \sum_{u'_e\! u_e} \sum_{u'_{\rm sto}\! u_{\rm sto}}\!\!\! |\bar{u}'_e(p'_e) \gamma_\nu u_e(p_e)\frac{1}{k^2}\bar{u}'_{\rm sto}(p'_{\rm sto}) \!\left[2m_{\rm sto}\gamma^\nu \!-\! (p'_{\rm sto} \!+\! p_{\rm sto})^\nu  \right]\! u_{\rm sto}(p_{\rm sto})|^2 , ({\rm B}5)
\end{eqnarray*}


\ni where we get

$$
m^2_e \sum_{u'_e u_e}\!\sum_{u'_{\rm sto} u_{\rm sto}} |\;\;|^2 = 4\left(1 - 2\frac{p_e \cdot p_{\rm sto}}{m^2_{\rm sto}}\right) + 16\left[ m^2_e - \left(\frac{p_e \cdot p_{\rm sto}}{m_{\rm sto}}\right)^2\right]\frac{1}{k^2} \;,
\eqno{\rm (B6)}
$$

\ni evaluating traces in Dirac bispinor indices. Here, in the sterino rest frame, where $\vec{p}_{\rm sto} = 0$, the collision relative velocity is $v_{\rm flux} = v_e  = |\vec{p}_e|/E_e$. In this Appendix, $ \sigma$ denotes $\sigma(e^-\,{\rm sto} \rightarrow e^-\,{\rm sto})$. 

Finally, we can estimate the electron differential cross-section on sterinos:

\begin{eqnarray*}
\frac{d \sigma}{d \Omega_e} \!\!\!& \!\!=\!\! &  \int_{0}^{\infty}{{\vec{p}\,'\!\!}_e}^2 d|{\vec{p}\,'\!\!}_e|\,\int d^3 {\vec{p}\,'\!\!}_{\rm sto}\frac{d^6 \sigma}{d^3 {\vec{p}\,'\!\!}_e\, d^3 {\vec{p}\,'\!\!}_{\rm sto}} \\
&\!\!=\!\! & \frac{1}{v_{\rm flux}}\left( \frac{2e \zeta f <\varphi_{\rm stn}>_{\rm vac}}
{\pi M^2}\right)^2 \!\frac{{\vec{p}\,'\!\!_e}^2}{ {(E_e + E'_{\rm sto}})|{\vec{p}\,'\!\!}_e| - E'_e|{\vec{p}}_e|\cos \theta_e}\; \frac{m^2_{\rm sto}}{E_e E_{\rm sto}} \\
& \!\!\times\!\! & \left\{1-2\frac{p_e \cdot p_{\rm sto}}{m^2_{\rm sto}} + 4\left[ m^2_e - \left(
\frac{p_e \cdot p_{\rm sto}}{m_{\rm sto}} \right)^2\right]\frac{1}{k^2}\right\}\;,\hspace{5.5cm} ({\rm B}7)
\end{eqnarray*}

\ni where $d\Omega_e = 2\pi \sin \theta_e d\theta_e$ and $\cos \theta_e = {\vec{p}\,'\!\!}_e\cdot 
\vec{p}_e/(|{\vec{p}\,'\!\!}_e||\vec{p}_e|)$. Here, ${\vec{p}\,'\!\!}_e + {\vec{p}\,'\!\!}_{\rm sto} = p_e + p_{\rm sto}$ giving for $k = {\vec{p}_e - \vec{p}\,'\!\!}_e$ the first of relations

$$
k^2 = -2 k\cdot p_{\rm sto}\;,\; k^2 = 2\left(m^2_e -E'_e E_e + |{\vec{p}\,'\!\!}_e||\vec{p}_e|\cos \theta_e \right)\,, \eqno{\rm (B8)}
$$

\ni the second following from the definition of $k$.

In the sterino rest frame, where $\vec{p}_{\rm sto} = 0$,  Eq. (B7) takes the form 

\begin{eqnarray*}
\hspace{2,0cm}\frac{d \sigma}{d \Omega_e} \!\!\!& \!\!=\!\! & \left( \frac{2e \zeta f <\varphi_{\rm stn}>_{\rm vac}} {\pi M^2}\right)^2 \frac{|{\vec{p}\,'\!\!}_e|}{|\vec{p}_e|} \;\frac{m_{\rm sto}}{E_e + m_{\rm sto} - (E'_e |\vec{p}_e|/|{\vec{p}\,'\!\!}_e|)\cos \theta_e} \\
& \!\!\times\!\! & \left( 1-2\frac{E_e }{m_{\rm sto}} + 4\frac{ m^2_e - E^2_e}{k^2} \right)\;.\hspace{6.6cm} ({\rm B}9)
\end{eqnarray*}

\ni Here, the first relation (B8) gives

$$
k^2 = 2\left(E'_e - E_e \right) m_{\rm sto} 
\eqno{\rm (B10)}
$$

\ni and then, together with the second relation (B8) implies

$$
(E_e + m_{\rm sto}) E'_e - |\vec{p}\,'\!\!_e| |\vec{p}_e| \cos\theta_e  = m^2_e + E_e m_{\rm sto} \;.
\eqno{\rm (B11)}
$$

If $ m_e/E_e \ll 1$ {\it i.e.}, the electron mass is negligible, then from the second Eq. (B8)

$$
k^2 \simeq -4 E'_e E_e \sin^2 \frac{\theta_e}{2} 
\eqno{\rm (B12)}
$$

\ni and Eq. (B11) gives

$$
\frac{E_e}{ E'_e} \simeq 1 + \frac{2E_e }{m_{\rm sto}} \sin^2 \frac{\theta_e}{2}  \,,
\eqno{\rm (B13)}
$$ 

\ni while the denominator in Eq. (B9) becomes 

$$
E_e + m_{\rm sto} - (E'_e |\vec{p}_e|/|{\vec{p}\,'\!\!}_e|) \cos \theta_e \simeq m_{\rm sto}\left(1 + \frac{2E_e }{m_{\rm sto}}\sin^2 \frac{\theta_e}{2}\right) \simeq \frac{E_e m_{\rm sto}}{E'_e}  \;.
\eqno{\rm (B14)}
$$

\ni Thus, if $ m_e/E_e \ll 1$, we get from Eq. (B9)
 
$$
\frac{d \sigma}{d \Omega_e} \simeq \left( \frac{2e \zeta f <\varphi_{\rm stn}>_{\rm vac}} {\pi M^2}\right)^2 \frac{1 + \sin^2 \frac{\theta_e}{2}}{\sin^2 \frac{\theta_e}{2}\left[1+(2E_e/m_{\rm sto})\sin^2 \frac{\theta_e}{2}\right]}\;.
\eqno{\rm (B15)}
$$

\ni We can see that for our quasi-electromagnetic interactions between electrons and sterinos the forward singularity still appears, though it is softer than for the \SM electromagnetic interaction of electrons and, say, point-like protons, where the differential electron cross-section on protons takes the form [11]

$$
\frac{d \sigma}{d \Omega_e} \simeq \left( \frac{e^2} {4\pi }\right)^2\; \frac{1}{4E^2_e} \; \frac{\cos^2 \frac{\theta_e}{2} - (k^2/2m_p)\sin^2 \frac{\theta_e}{2}}{\sin^4 \frac{\theta_e}{2}\left[1+(2E_e/m_p)\sin^2 \frac{\theta_e}{2}\right]}\;,
\eqno{\rm (B16)}
$$

\ni valid if $ m_e/E_e \ll 1$ in the proton rest frame. Here,

$$
k^2 \simeq -4 E'_e E_e \sin^2 \frac{\theta_e}{2}\simeq -4 E^2_e \,\frac{\sin^2 \frac{\theta_e}{2}}{1 + (2E_e/m_p) \sin^2 \frac{\theta_e}{2}}\;.
\eqno{\rm (B17)}
$$

On the contrary, if $ E_e/m_{\rm sto} \ll 1$ in the sterino rest frame {\it i.e.}, the sterino recoils are negligible, then $|{\vec{p}\,'\!\!}_e| \simeq |\vec{p}_e|\,,\; E'_{\rm sto} \simeq E_{\rm sto} = m_{\rm sto}$ and so, Eq. (B7) gives

$$
\frac{d \sigma}{d \Omega_e} \simeq\left( \frac{2e \zeta f <\varphi_{\rm stn}>_{\rm vac}} {\pi M^2}\right)^2 \frac{1 + \sin^2\frac{\theta_e}{2}}{\sin^2\frac{\theta_e}{2} } \,,
\eqno{\rm (B18)}
$$

\ni since in this case

$$
k^2 = (p_e - p'_e)^2 \simeq 2(m^2_e - E^2_e  + \vec{p}^{\,2}_e \cos^2 \theta_e) = -4 \vec{p}^{\,2}_e \sin^2 \frac{\theta_e}{2}\;.
\eqno{\rm (B19)}
$$

\ni The forward singularity in Eq. (B18) is softer than in the \SM electron differential cross-section on, say, point-like protons, valid if $ E_e/m_p \ll 1$ in the proton rest frame (Mott cross-section). This is of the form [11]

$$
\frac{d \sigma}{d \Omega_e} \simeq \left( \frac{e^2} {4\pi }\right)^2\; \frac{4E^2_e}{4{\vec{p}^{\,4}_{e}}} \; \frac{1 - (\vec{p}^{\,2}_e/E^2_e)\sin^2 \frac{\theta_e}{2}}{\sin^4 \frac{\theta_e}{2}}
\eqno{\rm (B20)}
$$

\ni with $|{\vec{p}\,'\!\!}_e| \simeq |\vec{p}_e|$ and $k^2 \simeq -4 \vec{p}^{\,2}_e \sin^2 (\theta/2)$. 

In contrast to the elastic scattering of electrons on sterinos ($e^-({\rm sterino}) \!\rightarrow\!$ $e^-{\rm (sterino)}$), for the crossed process (18) of annihilation of a sterino--antisterino pair into an electron--positron pair ((antisterino)(sterino) $\rightarrow e^+ e^-$) the corresponding differential cross-section $ d\sigma({\rm asto\,sto} \rightarrow e^+ e^-)/d\Omega_{e^+}$ calculated in our photonic portal can be integrated. Then, in the sterino--antisterino centre-of-mass frame, where the relative velocity of the colliding sterino--antisterino pair is $2v_{\rm sto}$, we obtain the following formula for total cross-section multiplied by sterino velocity $v_{\rm sto} = |\vec{p}_{\rm sto}|/E_{\rm sto} = \sqrt{1-m^2_{\rm sto}/E^2_{\rm sto}}$:

$$
\sigma({\rm asto\,sto} \rightarrow e^+ e^-)\, v_{\rm sto} \simeq  \frac{e^2} {4\pi }\,\left(\frac{2 \zeta f <\varphi_{\rm stn}>_{\rm vac}} { M^2} \right)^2\frac{8}{3}\, \frac{E^2_{\rm sto}+2m^2_{\rm sto}}{E^2_{\rm sto}} \;,
\eqno{\rm (B21)}
$$

\ni if $ m_e/E_{\rm sto} \ll 1$ {\it i.e.}, the electron mass is negligible.

However, the simplest annihilation channel of a sterino--antisterino pair is that leading into a photon and a physical steron (see process (26)). The corresponding total cross-section multiplied by the sterino velocity gets in our photonic portal the following form:

$$
\sigma({\rm asto\,sto} \rightarrow \gamma \;{\rm stn}) \,v_{\rm sto} = \frac{1} {4\pi } \left(\frac{2\zeta f}{M^2} \right)^2\; \frac{4}{3}\left(E^2_{\rm sto} + 2m^2_{\rm sto}\right)\left(1 - \frac{m^2_{\rm stn}}{4E^2_{\rm sto}} \right)
\eqno{\rm (B22)}
$$

\ni in the sterino--antisterino centre-of-mass frame (of course, if masses allow {\it i.e.}, if $ m_{\rm stn}< 
2m_{\rm sto} \leq 2E_{\rm sto}$ or $2 m_{\rm sto}\leq m_{\rm stn} < 2E_{\rm sto}$ for a given sterino energy $E_{\rm sto}$; in the second option, with $ m_{\rm stn}$ large enough, the channel (26), (antisterino)(sterino) $\rightarrow \gamma$(steron), may be closed for realistic sterino energies $E_{\rm sto}$).

\vfill\eject

~~~~

{\centerline{\bf References}}

\vspace{0.2cm}

\baselineskip 0.72cm

{\everypar={\hangindent=0.65truecm}
\parindent=0pt\frenchspacing

{\everypar={\hangindent=0.65truecm}
\parindent=0pt\frenchspacing

~[1]~For recent comments {\it cf.} M.E.~Peskin, arXiv: 0707.1536 [{\tt hep-ph}]; F.~Wilczek, arXiv: 0708.4036 [{\tt hep-ph}]. 

\vspace{0.1cm}

~[2]~For a recent review {\it cf.} R.~Essig, arXiv: 0710.1668 [{\tt hep-ph}].

\vspace{0.1cm}

~[3]~For a recent review {\it cf.} D.~Hooper, arXiv: 0710.2062 [{\tt hep-ph}].

\vspace{0.1cm}

~[4]~J.J.~van der Bij, {\it Phys. Lett.} {\bf B 636}, 56 (2006); arXiv: 0707.0359 [{\tt hep-ph}]; D~O'Connell, M.J.~Ramsay-Musolf and M.B.~Wise, {\it Phys. Rev.} {\bf D 75}, 037701 (2007); and references therein. 

\vspace{0.1cm}

~[5] For a recent presentation {\it cf.}\, W. Kr\'{o}likowski, {\it Acta Phys. Polon.} {\bf B 38} (to appear); 
W.~Kr\'{o}likowski,  {\tt hep--ph/0604148} and {\tt hep--ph/0504256}; and references therein. 

\vspace{0.1cm}
 
~[6]~P. Minkowski, {\it Phys. Lett.} {\bf B 67}, 421 (1977); M.~Gell-Mann, P.~Ramond and R.~Slansky, in {\it Supergravity}, ed. by F.~van~Nieuvenhuizen and D.~Freedman (North Holland, Amsterdam, 1979); T.~Yanagida, in {\it Proc. of the Workshop on Unified Theory and the Baryon Number of the Universe}, ed. by O.~Sawada and A.~Sugamoto (KEK, Sukuba, 1979); S.L.~Glashow, in {\it Quarks and Leptons}, ed. by 
M.~L\'{e}vy {\it et al.} (Plenum, New York, 1980); R.N.~Mohapatra and G.~Sanjanovic, {\it Phys. Rev. Lett.} {\bf 44}, 912 (1980). 

\vspace{0.1cm}

~[7]~R. Schabinger and J.D.~Wells, {\it Phys. Rev.} {\bf D 72}, 093007 (2005); B.~Patt and F.~Wilczek, 
{\tt hep--ph/0605188}; O.~Bartolomi and R.~Rosenfeld, arXiv: 0708.1784 [{\tt hep-ph}]. 

\vspace{0.1cm}

~[8]~For a recent discussion {\it cf.}\, Y.~Ascasibar, P.~Jean, C.~Boehm and J.~Kn\"{o}dlseder, {\tt \,astro-ph/0507142};  C.~Boehm and J.~Silk, arXiv: 0708.2768 [{\tt hep-ph}];  and references therein. 

\vspace{0.1cm}

~[9]~P. Jordan, {\it Schwerkraft und Weltall} (Vieweg, Braunschweig,1955); M.~Fierz, {\it Helv. Phys. Acta} {\bf 29}, 128 (1956); C.~Brans and R.H.~Dicke, {\it Phys. Rev.} {\bf 124}, 925 (1961); T.~Damour and K.~Nordtvadt, {\it Phys. Rev. Lett.} {\bf 70}, 2217 (1993); R.~Catena, M.~Pietroni  and L.~Scarabello,
 {\tt astro--ph/0604492}.

\vspace{0.1cm}

[10]~X.~Calmet, arXiv: 0708.2767 [{\tt hep-ph}]; R.~Catena and J.~M\"{o}ller, arXiv: 0709.1931 [{\tt hep-ph}]. 

\vspace{0.1cm}
 
[11]~J.D.~Bjorken and S.D.~Drell, {\it Relativistic Quantum Mechanics} (McGraw-Hill, New York, 1964). 

\vfill\eject

\end{document}